# Ultrafast Inorganic Crystals with Mass Production Capability for Future High-Rate Experiments

Chen Hu, Liyuan Zhang, and Ren-Yuan Zhu*[a]
[a] 256-48, HEP, California Institute of Technology, Pasadena, CA 91125, USA

## ABSTRACT

Future HEP experiments present stringent challenges to inorganic scintillators in both fast timing response and radiation tolerance. This paper reports recent progress in developing ultrafast inorganic scintillators with sub-ns decay time for future precision timing detectors and high-rate experiments. Performance of fast and ultrafast crystals with mass production capability are compared to CsI crystals which are used for the Mu2e calorimeter. Examples are LYSO:Ce, $BaF_2$ and $BaF_2$:Y, which are considered for Mu2e-II. Crystal radiation hardness against gamma-rays and hadrons is reported. Current status and development effort for the $BaF_2$:Y crystals are discussed.

## INTRODUCTION

Inorganic scintillators are widely used in HEP experiments to construct electromagnetic calorimeters, providing the best possible energy resolution and position resolution, good electron and photon identification and reconstruction efficiency. The recent DOE report on basic research needs for HEP instruments [1] points out that ultrafast inorganic scintillators with good radiation hardness are required for future HEP experiments at the energy and intensity frontiers to mitigate severe radiation environment up to 100 Mrad and $3\times10^{16}$ $n_{eq}/cm^2$ of one MeV equivalent neutron fluence [2] and high event rate and pileup [3], respectively. Development of ultrafast heavy crystals with sub-nanosecond decay time thus is important to break the ps timing barrier for time of fight (TOF) systems and for ultrafast calorimetry. Table 1 lists optical and scintillation properties for some fast and ultrafast inorganic scintillators with mass production capability. Also listed in Table 1 is the figures of merit for the TOF application, which is the light yield (LY) in the 1st ns and the ratio between the LY in the 1st nanosecond and the total LY.

**Table 1** Optical and scintillation properties of fast and ultrafast inorganic scintillators with mass-production capability

| | CsI | LYSO:Ce | $BaF_2$ | $BaF_2$:Y | $CeBr_3$ | $LaCl_3$ | $LaBr_3$ |
|---|---|---|---|---|---|---|---|
| Density (g/cm$^3$) | 4.51 | 7.4 | 4.89 | 4.89 | 5.23 | 3.86 | 5.29 |
| Melting points (°C) | 621 | 2050 | 1280 | 1280 | 722 | 858 | 783 |
| $X_0$ (cm) | 1.86 | 1.14 | 2.03 | 2.03 | 1.88 | 2.81 | 1.88 |
| $R_M$ (cm) | 3.57 | 2.07 | 3.10 | 3.10 | 2.88 | 3.71 | 2.85 |
| $\lambda_I$ (cm) | 39.3 | 20.9 | 30.7 | 30.7 | 30.8 | 37.6 | 30.4 |
| $Z_{eff}$ | 54.0 | 64.8 | 51.6 | 51.6 | 46.2 | 47.3 | 45.6 |
| dE/dX (MeV/cm) | 5.56 | 9.55 | 6.52 | 6.52 | 6.81 | 5.27 | 6.90 |
| $\lambda_{peak}$ [a] (nm) | 420 / 310 | 420 | 300 / 220 | 300 / 220 | 371 | 335 | 356 |
| Refractive Index[b] | 1.95 | 1.82 | 1.50 | 1.50 | 1.9 | 1.9 | 1.9 |
| Normalized Light Yield[a,c] | 4.2 / 1.3 | 100 | 42 / 4.8 | 1.7 / 4.8 | 99 | 15 / 49 | 153 |
| Total Light yield (ph/MeV) | 1,650 | 30,000 | 13,000 | 2,000 | 30,000 | 19,000 | 46,000 |
| Decay time[a] (ns) | 30 / 6 | 40 | 600 / 0.5 | 600 / 0.5 | 17 | 570 / 24 | 20 |
| LY in 1st ns (photons/MeV) | 100 | 740 | 1,200 | 1,200 | 1,700 | 610 | 1,500 |
| 1st ns LY/Total LY (%) | 6.1% | 2.5% | 9.2% | 60% | 5.7% | 3.2% | 3.3% |
| Comment | Mu2e-II Candidate | Mu2e-II Candidate | Mu2e-II Candidate | Mu2e-II Candidate | hygroscopic | hygroscopic | hygroscopic |

[a] top/bottom row: slow/fast component; [b] at the emission peak; [c] normalized to LYSO:Ce.

*zhu@caltech.edu; phone 1 626 395-6661; fax 1 (626) 395-8728; http://www.hep.caltech.edu/~zhu/

Cerium doped lutetium yttrium oxyorthosilicate ($Lu_{2(1-x)}Y_{2x}SiO_5$:Ce or LYSO:Ce) shows high stopping power, high light output, fast decay time and good radiation hardness against both ionization dose and hadrons [4]. LYSO:Ce thus is now used to construct the Barrel Timing Layer (BTL) for the CMS experiment at the HL-LHC and was one of the original choices of the Mu2e experiment at Fermilab. The high cost of LYSO:Ce caused by raw material $Lu_2O_3$ and high melting point, however, limited its use. The Mu2e experiment is building an undoped CsI total absorption calorimeter [5] and considers an ultrafast total absorption calorimeter for its upgrade [3]. Inorganic scintillator with core valence transition, such as $BaF_2$, features with its energy gap between the valence band and the uppermost core band less than the fundamental bandgap, allowing an ultrafast decay time. By using its ultrafast scintillation component with 0.5 ns decay time, a $BaF_2$ calorimeter promises an ultrafast calorimeter for future HEP experiments. Yttrium doped barium fluoride ($BaF_2$:Y) shows a significantly reduced slow component and radiation-induced readout noise [4], so is promising for Mu2e-II. R&D is on-going in collaboration with crystal producers to develop $BaF_2$:Y crystals of large size [6]. Other bright and fast crystals, such as $CsF$, $CeBr_3$, $LaCl_3$:Ce and $LaBr_3$:Ce, are hygroscopic, presenting a technical challenge for calorimeter construction. It is also interesting to note that ultrafast inorganic scintillators listed in Table 1 may also be used beyond HEP for e.g., GHz hard X-ray imaging for future free electron laser facilities [7].

## RADIATION HARDNESS

Radiation hardness against gamma-rays was measured for various crystals of large size up to 340 Mrad [8]. Fig. 1 shows the normalized emission weighted longitudinal transmittance (EWLT, top) and the normalized light output (LO, bottom) as a function of the integrated dose for four CsI (left), six LYSO/LSO/LFS (middle) and three $BaF_2$ (right) crystals of large size from various vendors. The EWLT was calculated according to

$$EWLT = \frac{\int T(\lambda)Em(\lambda)d\lambda}{\int Em(\lambda)d\lambda} \qquad (1)$$

where $T(\lambda)$ and $Em(\lambda)$ are transmittance and emission spectra. The EWLT value provides a numerical representation of the transmittance over the entire emission spectrum.

The results show that CsI survives the Mu2e radiation environment where an ionization dose up to a few tens krad, but not 1 Mrad expected by Mu2e-II. Improving on the decay time and radiation hardness of pure CsI is necessary to meet the more stringent requirements of Mu2e-II. Undoped $BaF_2$ crystals show saturated damage from 10 krad to 100 Mrad after an initial loss, so are more radiation hard than CsI at a large integrated dose. In conclusion, LYSO:Ce and $BaF_2$ survive the radiation environment expected by Mu2e-II.

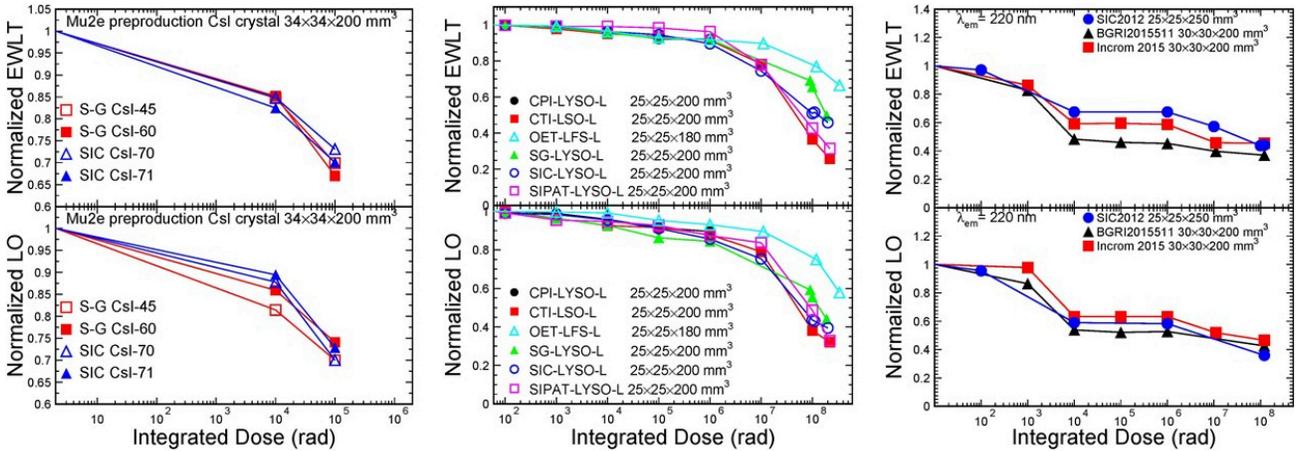

**Figure 1.** The normalized EWLT (top) and light output (LO) are shown as a function of the integrated dose for four CsI (left), six LYSO/LSO/LFS (middle) and three $BaF_2$ (right) crystals from various vendors.

Figure 2 shows transmittance spectra (left), light output as a function of integration time (middle) and normalized light output (right) as a function of 800 MeV proton fluence for six $BaF_2$ plates of $25\times25\times5$ mm$^3$ and compared to six LYSO:Ce plates of 10×10×3 mm³ and six PWO plates of 25×25×5 mm³. Both LYSO and BaF₂ plates show more than 80% of light output after $10^{16}$ p/cm²

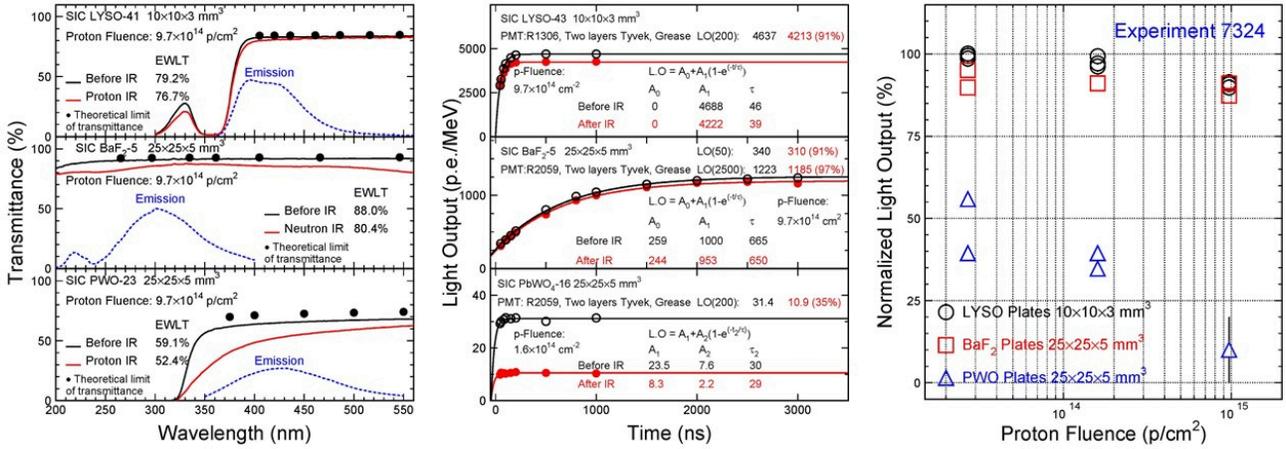

**Figure 2.** Transmittance spectra (left) and light output as a function of integration time (middle) and light output (right) are shown as a function of 800 MeV proton fluence for six BaF$_2$ plates of 25×25×5 mm³ and compared to six LYSO:Ce plates of 10×10×3 mm³ and six PWO plates of 25×25×5 mm³.

Figure 3 shows transmittance spectra (left), light output as a function of integration time (middle) and normalized light output (right) as a function of one MeV equivalent neutron fluence for six BaF$_2$ plates of 15×15×5 mm³ and compared to six LYSO:Ce plates of 10×10×5 mm³ and six PWO plates of 15×15×5 mm³. Both LYSO:Ce and BaF$_2$ plates show more than 80% of light output after $10^{16}$ n$_{eq}$/cm².

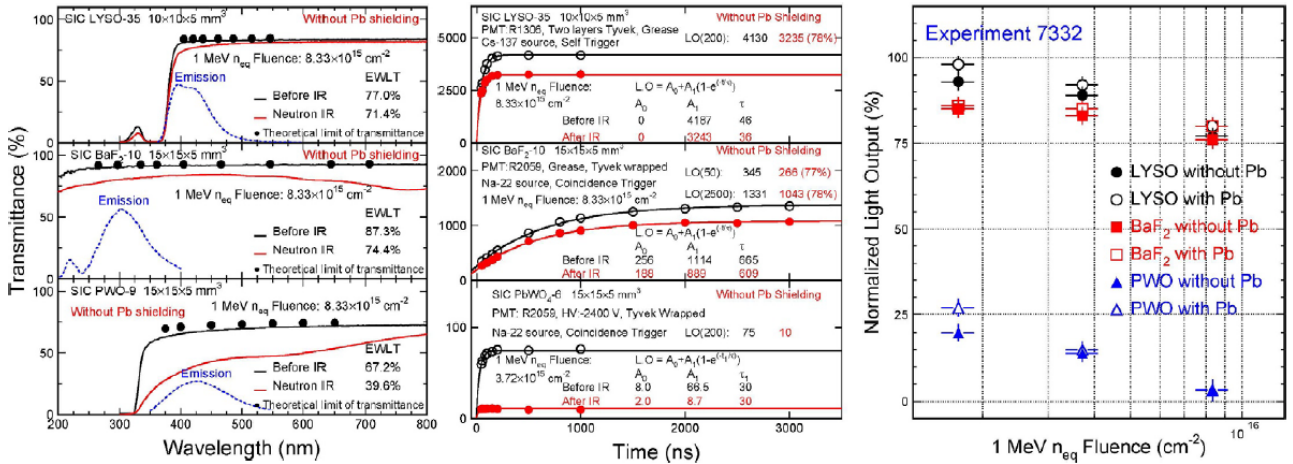

**Figure 3.** Transmittance spectra (left) and light output as a function of integration time (middle) and light output (right) are shown as a function of one MeV equivalent neutron fluence for six BaF$_2$ plates of 15×15×5 mm³ and compared to six LYSO:Ce plates of 10×10×5 mm³ and six PWO plates of 15×15×5 mm³.

## PROGRESS ON ULTRAFAST BaF$_2$:Y CRYSTALS

It is well known that BaF$_2$ crystals have an ultrafast cross-luminescence scintillation with sub-ns decay time peaked at 220 nm, and a 600 ns slow component peaked at 300 nm with a much higher intensity. The latter causes pileup in a high-rate environment. The left plot of Fig. 4 shows the pulse shape measured by a PMT (top) and a MCP (bottom) for a BaF$_2$ sample. About 0.5 and 0.9 ns of decay time and FWHM width respectively are observed by the MCP, but not PMT, where 1.4 and 3.1 ns are observed due to slow response time of the PMT. It is also known that the slow component in BaF$_2$ crystals may be suppressed either by rare earth doping in crystals [4] or by using a solar blind photodetector [9]. The middle and right plots of Fig. 4 are respectively the X-ray excited emission spectra and the light output as a function of integration time for BaF$_2$ cylinders of Φ18×21 mm$^3$ grown at Beijing Glass Research Institute (BGRI) with different Y$^{3+}$ doping levels [4]. They show a reduced slow light intensity for an increased yttrium doping level, while the intensity of the fast emission is maintained.

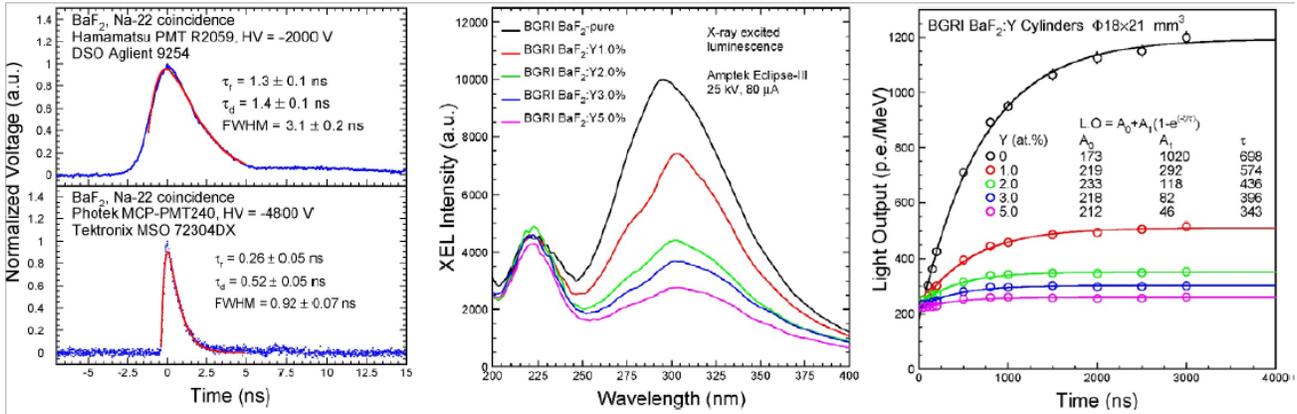

**Figure 4.** Left: The pulse shape measured by a PMT (top) and a MCP (bottom) shows the ultrafast scintillation light component with 0.5 ns decay time for a BaF$_2$ sample. The emission (Middle) and integrated light output (Right) are shown for BaF$_2$ samples with yttrium doping at different levels.

Development for large size BaF$_2$:Y crystals with better optical quality, scintillation performance, and radiation hardness is on-going [6]. The left plot of Fig. 6 shows the transmittance spectra for two BaF$_2$:Y crystals grown recently by BGRI (top) and SIC (bottom) together with BaF$_2$:Y emission spectrum and the numerical values of EWLT [11]. Progress was observed in optical quality. The middle plot of Fig. 6 shows photocurrent measured by the Hamamatsu R2059 PMT for SIC BaF$_2$:Y-2020 (top) and SIC BaF$_2$-2 (bottom) under 23 rad/h, respectively. The right plot of Fig. 6 shows the measured photocurrent as a function of the dose rate for BGRI and SIC BaF$_2$:Y and SIC BaF$_2$-2 samples under 2 and 23 rad/h. BaF$_2$:Y long crystals with excellent optical quality and much-reduced impurity related absorption were fabricated successfully. Yttrium doping in BaF$_2$ reduces the γ-ray induced photocurrent and readout noise significantly. R&D is on-going to measure the radiation hardness for the BaF$_2$:Y crystals.

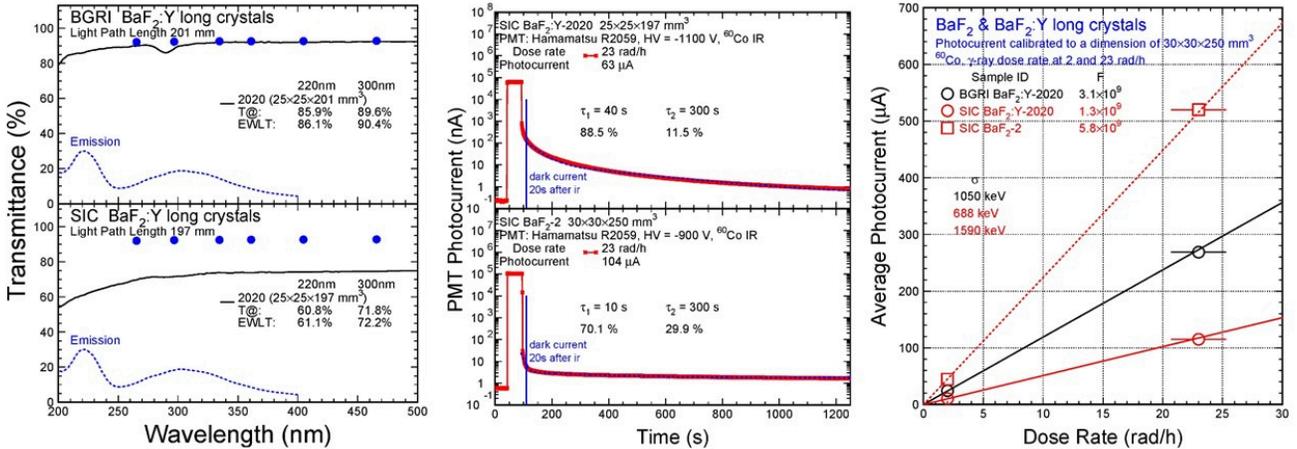

**Figure 5.** Left: Transmittance spectra for two BaF$_2$:Y crystals from BGRI (top) and SIC (bottom). Middle: Photocurrent measured by the Hamamatsu R2059 PMT for SIC BaF$_2$:Y-2020 (top) and SIC BaF$_2$-2 (bottom) under 23 rad/h, respectively. Right: photocurrent is shown as a function of the dose rate for BGRI and SIC BaF$_2$:Y samples and SIC BaF$_2$-2 under 2 and 23 rad/h.

In addition to yttrium doping in BaF$_2$ crystals solar-blind photodetectors also improve the F/S ratio, and thus reduce radiation induced readout noise to a level of less than 1 MeV for a BaF$_2$:Y crystal-based ultrafast calorimeter [12]. Fig. 7 shows the quantum efficiency (QE) for a Photek solar-blind cathode (Left), the photon detection efficiency (PDE) for a

FBK solar-blind SiPM (Middle) and the PDE of a Hamamatsu VUV SiPM (Right) as a function of wavelength and the emission spectra of BaF$_2$ and BaF$_2$:Y [12]. Also shown in the figures are the emission spectra of BaF$_2$ and BaF$_2$:Y crystals, and the numerical values of the emission weighted QE (EWQE) and PDE (EWPDE) for the fast and slow components. The EWQE and EWPDE values represent photodetector's ability to detect the fast and slow component. Their ratio (F/S) represents photodetector's ability for the slow suppression. Both solar-blind photodetectors show higher fast detection efficiency and larger F/S ration as compared to the Hamamatsu VUV SiPM. R&D continues along this line to develop solar-blind photodetectors.

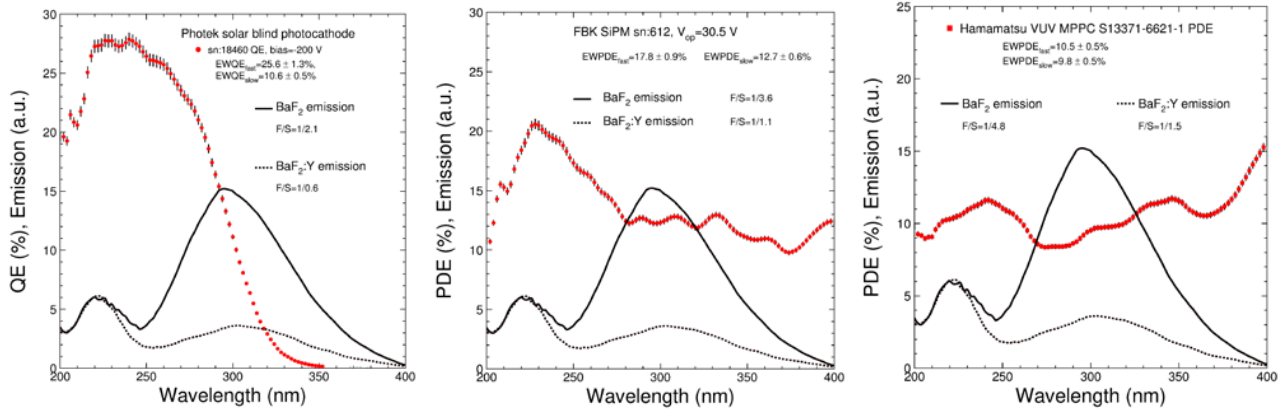

**Figure 6.** The QE of a Photek solar-blind cathode (Left), the PDE of a FBK solar-blind SiPM (Middle) and the PDE of a Hamamatsu VUV SiPM (Right) are shown as a function of wavelength.

One of the potential applications of the ultrafast BaF$_2$:Y scintillation is front imager for GHz hard X-ray imaging required by future free electron laser facilities. Fig. 5 show response of BaF$_2$, BaF$_2$:Y, ZnO:Ga and LYSO:Ce crystals to septuplet X-ray bunches with 2.83 ns bunch spacing measured at the advanced photon source facility of ANL [10]. While BaF$_2$ crystals show clearly separated X-ray bunches with 2,83 ns spacing, the slow crystals do not. In addition, amplitude reduction is also observed for eight septuplets in BaF$_2$ and LYSO:Ce, which is due to MCP saturation caused by the slow scintillation in BaF$_2$ and LYSO:Ce, but not in slow-suppressed BaF$_2$:Y.

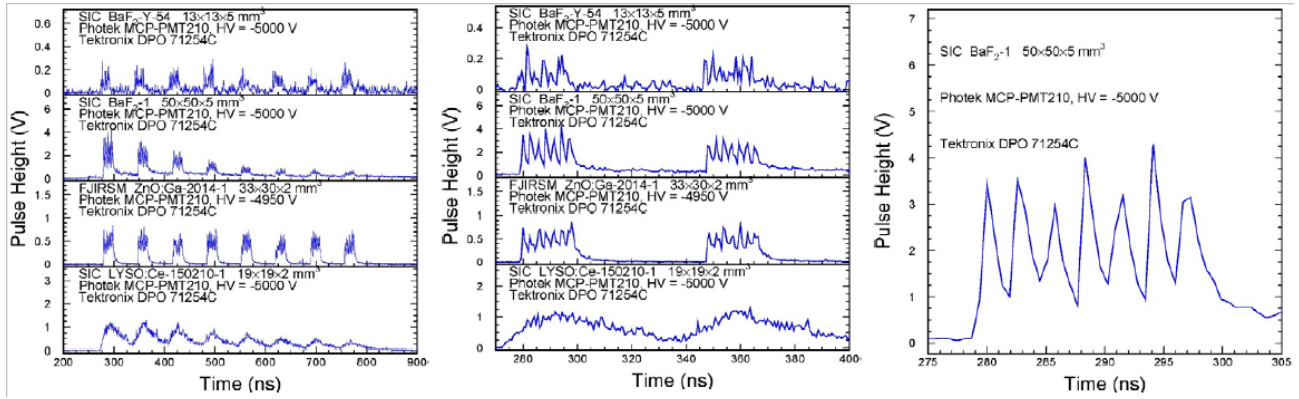

**Figure 7.** The temporal response to septuplet 30 keV X-ray bunches with 2.83 ns bunch spacing measured at APS of ANL is shown for BaF$_2$:Y, BaF$_2$, ZnO:Ga and LYSO:Ce crystal samples.

## SUMMARY

Future HEP experiments require fast and radiation hard inorganic scintillators. Scintillation performance of various fast and ultrafast crystals is summarized and compared. Bright, fast and radiation hard LYSO:Ce crystals meet the requirements but with high cost. Other crystals, such as CsF, CeBr$_3$, LaCl$_3$:Ce and LaBr$_3$:Ce, show excellent scintillation performance but are hygroscopic. Because of its ultrafast light of 0.5 ns decay time, $BaF_2$ crystals promise an ultrafast calorimeter for future HEP experiments. $BaF_2$:Y featured with the fast component and a suppressed slow component is also promising. R&D is on-going to develop large size $BaF_2$:Y crystals and solar-blind VUV photodetectors for Mu2e-II.

## ACKNOLEDGEMENTS

This work is supported by the U.S. Department of Energy, Office of High Energy Physics program under Award Number DE-SC0011925.